# Trapped Rainbow Techniques for Spectroscopy on a Chip and Fluorescence Enhancement

**Vera N. Smolyaninova [1], Igor I. Smolyaninov [2], Alexander V. Kildishev [3] and Vladimir M. Shalaev [3]**

[1] *Department of Physics Astronomy and Geosciences, Towson University, 8000 York Rd., Towson, MD 21252 USA*

*e-mail: vsmolyaninova@towson.edu  fax: +1-410-704-3511*

[2] *Department of Electrical and Computer Engineering, University of Maryland, College Park, MD 20742, USA*

[3] *Birck Nanotechnology Centre, School of Electrical and Computer Engineering, Purdue University, IN 47907, USA*

**We report on the experimental demonstration of the broadband "trapped rainbow" in the visible range using arrays of adiabatically tapered optical nano waveguides. Being a distinct case of the slow light phenomenon, the trapped rainbow effect could be applied to optical signal processing, and sensing in such applications as spectroscopy on a chip, and to providing enhanced light-matter interactions. As an example of the latter applications, we have fabricated a large area array of tapered nano-waveguides, which exhibit broadband "trapped rainbow" effect. Considerable fluorescence enhancement due to slow light behavior in the array has been observed.**

The concept of "trapped rainbow" has attracted considerable recent attention due to such potential applications as optical signal processing and spectroscopy on a chip, and due to expected enhancement of slow light interaction with matter. According to various theoretical models, a specially designed metamaterial [1] or plasmonic [2,3] waveguide has the ability to slow down and stop light of different wavelengths at different spatial locations along the waveguide. Unlike the typical narrow-band slow light schemes [4,5], the proposed theoretical trapped rainbow arrangements are extremely broadband, and can trap a true rainbow ranging from violet to red in the visible spectrum. However, due to the necessity of complicated nanofabrication, and the difficulty of producing broadband metamaterials, the trapped rainbow schemes had until recently remained in the theoretical domain only. Fortunately, very recently these roadblocks have been bypassed due to realization that the metamaterial properties necessary for transformation optics device fabrication can be emulated using various adiabatically tapered optical nano-waveguide geometries [6]. This realization has led to recent experimental demonstration of the broadband "trapped rainbow" in the visible frequency range [7].

In this paper we demonstrate an experimental realization of the broadband trapped rainbow effect which spans the 457-633 nm range of the visible spectrum. Similar to our recent demonstration of broadband cloaking [6], the metamaterial properties necessary for device fabrication were emulated using an adiabatically tapered optical nano waveguide geometry. Based on this initial demonstration, we have fabricated a large area array of tapered nano-waveguides, which exhibit broadband trapped rainbow effect. We have observed considerable fluorescence enhancement due to slow light behavior in the array. Since fluorescence enhancement strongly depends on the spatial location inside the waveguides, such "trapped rainbow arrays" can be used in multi-channel fluorescence sensors.

As a starting point of our experiments, we have demonstrated broadband trapped rainbow effect using a 4.5-mm diameter double convex glass lens which was coated on one side with a 30-nm gold film. The lens was placed with the gold-coated side down on top of a flat glass slide coated with a 70-nm gold film (Fig.1a). The air gap between these surfaces has been used as an adiabatically changing optical nano waveguide. The dispersion law of light in such a waveguide is

$$\frac{\omega^2}{c^2} = k_r^2 + \frac{k_\phi^2}{r^2} + \frac{\pi^2 l^2}{d(r)^2} \tag{1}$$

where $l = 1, 2, 3 \ldots$ is the transverse mode number, and $d(r)$ is the air gap, which is a function of radial coordinate $r$. Light from a multi-wavelength argon ion laser (operating at $\lambda = 457$ nm, 465 nm, 476 nm, 488 nm and 514 nm) and 633-nm light from a He-Ne laser was coupled to the waveguide via side illumination. This multi-line illumination produced the appearance of white light illuminating the waveguide. Light propagation through the nano waveguide was imaged from the top using an optical microscope (Fig. 1b). Since the waveguide width at the entrance point is large, the air gap waveguide starts as a multi-mode waveguide. Note that a photon launched into the $l$-th mode of the waveguide stays in this mode as long as $d$ changes adiabatically [8]. In addition, the angular momentum of the photons $k_\phi = \rho k = L$ is conserved (where $\rho$ is the impact parameter defined with respect to the origin). Gradual tapering of the waveguide leads to mode number reduction: similar to our observation of broadband cloaking (described in detail in ref. [6]) only $L=0$ modes may reach the vicinity of the point of contact between the gold-coated spherical and planar surfaces, and the group velocity of these modes

$$c_{gr} = c\sqrt{1 - \left(\frac{l\lambda}{2d}\right)^2} \qquad (2)$$

tends to zero as *d* is reduced: the colored rings around the central circular dark area in Fig.1c each represent a location where the group velocity of the *l*-th waveguide mode becomes zero. These locations are defined by

$$r_l = \sqrt{(l+1/2)R\lambda} \qquad (3)$$

where R is the lens radius [6]. Finally, the light in the waveguide is completely stopped at a distance

$$r = \sqrt{R\lambda/2} \qquad (4)$$

from the point of contact between the gold-coated surfaces, where the optical nano waveguide width reaches $d=\lambda/2\sim200$ nm range. The group velocity of the only remaining waveguide mode at this point is zero. This is consistent with the fact that the area around the point of contact appears dark in Fig. 1b. In this area the waveguide width falls below 200 nm down to zero. Since the stop radius depends on the light wavelength, different light colors stop at different locations inside the waveguide, which is quite obvious from Fig.1b. Thus, the visible light rainbow has been stopped and "trapped." This observation constitutes an experimental demonstration of a broadband trapped rainbow effect in the visible frequency range. Unlike recently described observation of the trapped rainbow in a left-handed heterostructure [9], the proposed geometries for trapping light in plasmonic waveguides [10], and controllable optical "black holes" [11], our geometry is easily scalable to any spectral range of interest. While the group velocity of the trapped photons is exactly zero at $r = \sqrt{R\lambda/2}$ (see eq.(2)), light cannot be "stored" indefinitely at these locations due to Joule losses in metal. In the best case scenario photons can be stored for no longer that 100-1000 periods. However, even this duration is enough to cause considerable enhancement of light-matter interaction in this geometry. Note that the light is stopped only for the waveguide mode, which has both the mode number *l=0* and the angular momentum number *L=0*. Therefore, these results do not contradict our observations of broadband cloaking [6]. In the ray optics approximation this condition corresponds to the central ray hitting the cloak. As was noted by Pendry *et al.*[12], such a ray "does not know" which way to turn around the cloak.

The described experimental arrangement may be used in such important applications as spectroscopy on a chip. Fig.2 presents comparison of the cross sections of the optical microscope images of the trapped rainbow effect obtained when all the spectral lines were present (Fig.2a) with the case when only two laser wavelengths (514 nm and 633-nm) are used for illumination (Fig.2b). Individual spectral lines separated by only a few micrometers appear to be well resolved. Based on the image cross section analysis, spectral resolution of the order of 40 nm has been obtained. Further improvement of spectral resolution may be achieved by using a gold-coated spherical surface with a smaller radius of curvature.

To further illustrate potential spectroscopic applications of this technique, we have fabricated a "multi-channel trapped rainbow" device and studied fluorescence enhancement due to slow light behavior. As a first fabrication step, a commercially available microlens array [13] (30 μm pitch, 42 μm lens radius) was coated on the microlens side with a 30-nm gold film (Fig.3a). The array was placed with the gold-coated side down on top of a flat glass slide coated with a 30-nm gold film. Periodic arrangement of nanometer scale gaps between these gold-coated surfaces has been used as an array of tapered optical nano waveguides. In our experiments the gap was filled with alcohol solutions of such widely used fluorescent dyes as fluorescein and rhodamine B.

Measured fluorescence spectra of these dyes are shown in Fig.3b. Similar to eq.(1), the dispersion law of light in the tapered region of each waveguide can be written as

$$\frac{\omega^2}{c^2} = k_r^2 + \frac{k_\phi^2}{r^2} + \frac{\pi^2 l^2}{n^2 d(r)^2} \tag{5}$$

where $l =1, 2, 3 \ldots$ is the transverse mode number, n is the solution refractive index, $d(r)$ is the gap between the gold-coated surfaces, which is a function of radial coordinate $r$, and $k_\phi=L$ is the angular momentum of the photons. The waveguide array was illuminated from the side using 488 nm light from an argon ion laser. Fluorescence excited in the waveguide array was imaged from the top using an optical microscope through the band-pass filter which cuts off 488 nm excitation light. Gradual tapering of each waveguide leads to mode number reduction: similar to our observations of the trapped rainbow, only *L=0* modes may reach the vicinity of the point of contact between the gold-coated spherical and planar surfaces, and the group velocity of these modes

$$c_{gr} = c\sqrt{1-\left(\frac{l\lambda}{2nd}\right)^2} \tag{6}$$

tends to zero as *d* is reduced. As a result, such "stopped light" interacts much stronger with the fluorescent dye molecules which is apparent from the microscopic images of dye fluorescence excited in the array of tapered waveguides (Fig.4). Since the stop radius depends on the light wavelength, different light colors stop at different locations inside the waveguide, which is quite obvious from Figs.4c,d obtained when both fluorescein and rhodamine B where present in the alcohol solution in the gap. The green and orange colored rings around the central circular dark area in Fig.4d each represent a location where the group velocity of the *l*-th waveguide mode becomes zero. These locations are defined by

$$r_l = \sqrt{(l+1/2)R\lambda/n} \qquad (7)$$

where R is the microsphere radius.

The increase in interaction time between light and the dye molecules (and the resulting fluorescence enhancement) can be estimated from the fluorescence linewidth Δλ of the dyes (Fig.3b). Assuming that photons at $\lambda=\lambda_0$ are stopped, and taking into account eq.(6), we obtain

$$\Delta c_{gr} = \left(\frac{2\Delta\lambda}{\lambda_0}\right)^{1/2} c \sim \frac{c}{3} \quad , \qquad (8)$$

resulting in the effective interaction time increase by a factor of 3. Quantitative analysis of the cross section (Fig.4b) of the fluorescence image indeed demonstrates approximately factor of 3 enhancement of fluorescence compared to the background fluorescence observed elsewhere in the waveguide. Note that this three-fold enhancement is observed on top of the enhancement due to waveguide geometry, which typically reaches factors of a few tens [14]. We should also note that the described "trapped rainbow array" scheme is capable of achieving much larger enhancement factors if the fluorescence linewidth Δλ is reduced. Together with spatial resolution demonstrated in Fig.4d, these observations indicate that the described experimental arrangement may be used in such important applications as spectroscopy on a chip. By design, our geometry is highly suitable for application in multi-channel fluorescence sensors and antibody arrays in which each tapered gap between the individual microsphere and the substrate in the array may act as a separate spatially selective fluorescence sensor.

In conclusion, we have reported experimental demonstration of the broadband "trapped rainbow" in the visible frequency range using an adiabatically tapered optical nano waveguide, and gave examples of its potential use in various spectroscopy on a chip applications. Being a distinct case of the slow light phenomenon, the trapped rainbow effect could be applied to optical computing and signal processing, and to providing enhanced light-matter interactions. A large area array of tapered nano-waveguides, which exhibit broadband "trapped rainbow" effect has been fabricated which is highly suitable for application in multi-channel fluorescence sensors and antibody arrays. Considerable fluorescence enhancement due to slow light behavior in the array has been observed.

**Acknowledgements**

V. Smolyaninova acknowledges support of this research by the NSF grants DMR-0348939 and DMR-1104676; V. Shalaev and A. Kildishev acknowledge support by ARO-MURI award 50342-PH-MUR

**REFERENCES**

1. K. L Tsakmakidis, A. D. Boardman, O. Hess, "Trapped rainbow storage of light in metamaterials", *Nature* **450**, 397 (2007).

2. M. I. Stockman, "Nanofocusing of optical energy in tapered plasmonic waveguides", *Phys. Rev. Letters* **93**, 137404 (2004).

3. Q. Gan, Y. J. Ding, F. J. Bartoli, "Rainbow trapping and releasing at telecommunication wavelengths", *Phys. Rev. Letters* **102**, 056801 (2009).

4. L. V. Hau, S. E. Harris, Z. Dutton, C. H. Behroozi, "Light speed reduction to 17 metres per second in an ultracold atomic gas", *Nature* **397**, 594 (1999).

5. Y. A. Vlasov, M. O'Boyle, H. F. Hamann, S. J. McNab, "Active control of slow light on a chip with photonic crystal waveguides", *Nature* **438**, 65 (2005).

6. I. I. Smolyaninov, V. N. Smolyaninova, A. V. Kildishev, and V. M. Shalaev, "Anisotropic metamaterials emulated by tapered waveguides: application to electromagnetic cloaking", *Phys. Rev. Letters* **103**, 213901 (2009).

7. V.N. Smolyaninova, I.I. Smolyaninov, A.V. Kildishev, and V.M. Shalaev, "Experimental observation of the trapped rainbow", *Appl. Phys. Letters* **96**, 211121 (2010).

8. L.D. Landau, E.M Lifshitz, Quantum Mechanics (Reed, Oxford, 1988).

9. X. P. Zhao, W. Luo, J. X. Huang, Q. H. Fu, K. Song, X. C. Cheng, and C. R. Luo, "Trapped rainbow effect in visible light left-handed heterostructures", *Appl. Phys.Lett.* **95**, 071111 (2009).

10. J. Park, K.-Y. Kim, I.-M. Lee, H. Na, S.-Y. Lee, and B. Lee, "Trapping light in plasmonic waveguides", *Opt. Express* **18**, 598 (2010).

11. Q. Bai, J. Chen, N.-H. Shen, C. Cheng, and H.-T. Wang, "Controllable optical black holes in left-handed materials", *Opt. Express* **18**, 2106 (2010).

12. J. B. Pendry, D. Schurig, and D. R. Smith, "Controlling electromagnetic fields", *Science* **312**, 1780 (2006).

13. http://www.suss-microoptics.com

14. A. Pokhriyal, M. Lu, C. S. Huang, S. Schulz, and B. T. Cunningham, "Multicolor fluorescence enhancement from a photonics crystal surface", *Appl. Phys. Letters* **97**, 121108 (2010).

**FIGURE CAPTIONS**

**Figure 1.** (a) Photo of the experimental geometry of the trapped rainbow experiment. A glass lens was coated on one side with a gold film. The lens was placed with the gold-coated side down on top of a flat glass slide also coated with a gold film. The air gap between these surfaces formed an adiabatically changing optical nano waveguide. HeNe and Ar:Ion laser light was coupled into the waveguide from the side. (b) Optical microscope image of the trapped rainbow.

**Figure 2.** Cross sections of the trapped rainbow images along the line shown in Fig.1b. Individual spectral lines are clearly resolved in the bottom plot obtained using 514 nm and 633 nm illumination. Multiple spectral lines are visible in the top cross section.

**Figure 3.** (a) Photo of the gold-coated microlens array (30 μm pitch, 42 μm lens radius) used in the "trapped rainbow array" fabrication. (b) Typical fluorescence spectra of alcohol solutions of fluorescein and rhodamine B.

**Figure 4.** (a) Image of fluorescence enhancement in the "trapped rainbow waveguide array" obtained when only fluorescein is present in the alcohol solution. (b) Image cross section along the red line shown in (a) indicates considerable fluorescence enhancement due to stopped light behavior: red arrows indicate locations of $c_{gr}$~0 for $l$=1,2 and 3 (c) Image of fluorescence enhancement in the "trapped rainbow waveguide array" obtained when both fluorescein and rhodamine B are present in the solution. (d) High resolution image of spatially resolved two dye fluorescence.

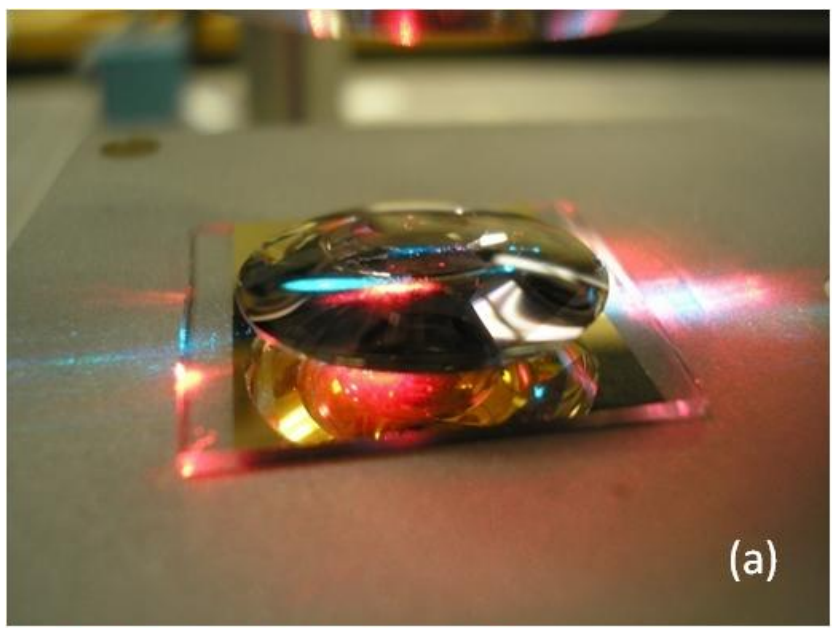

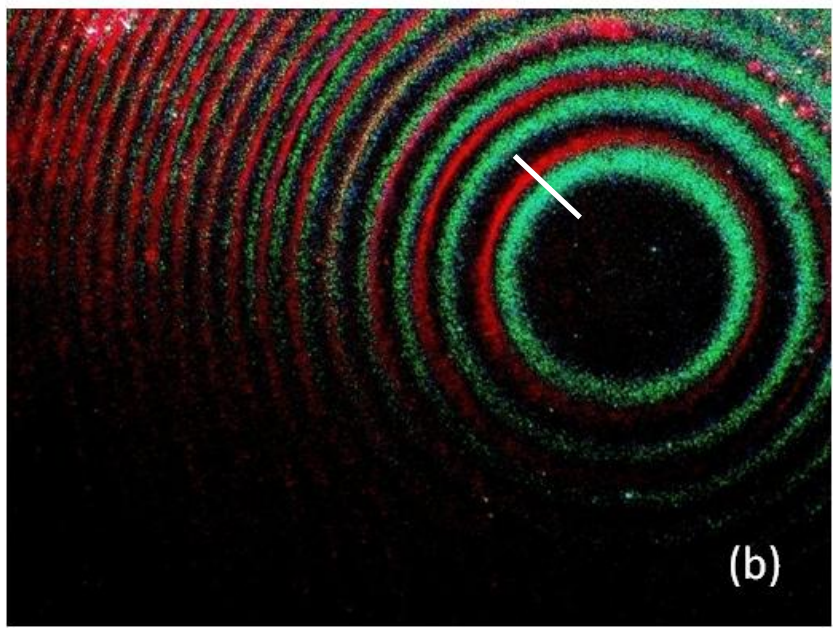

**Fig.1**

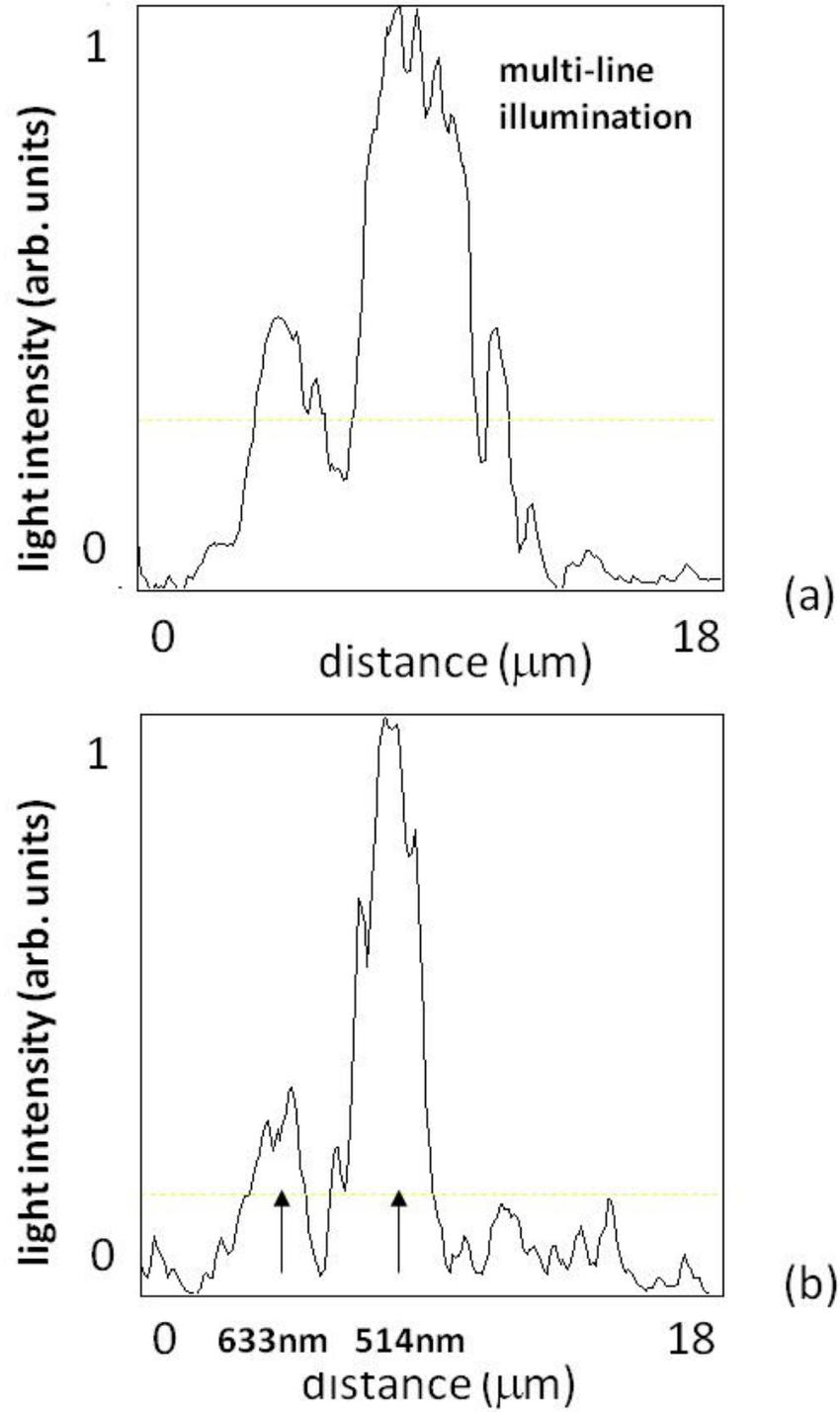

**Fig.2**

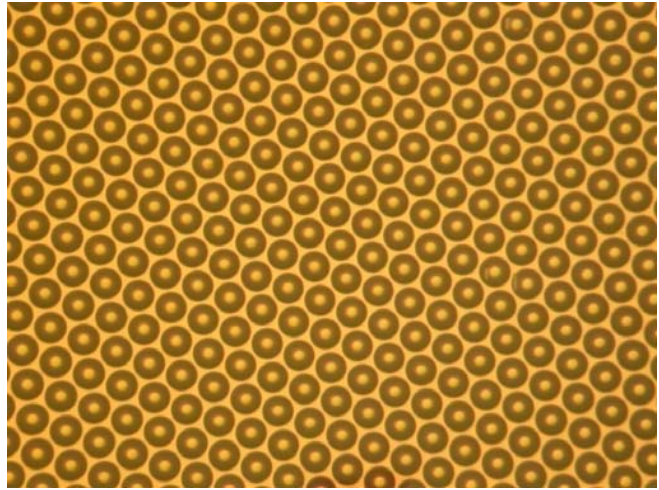

(a)

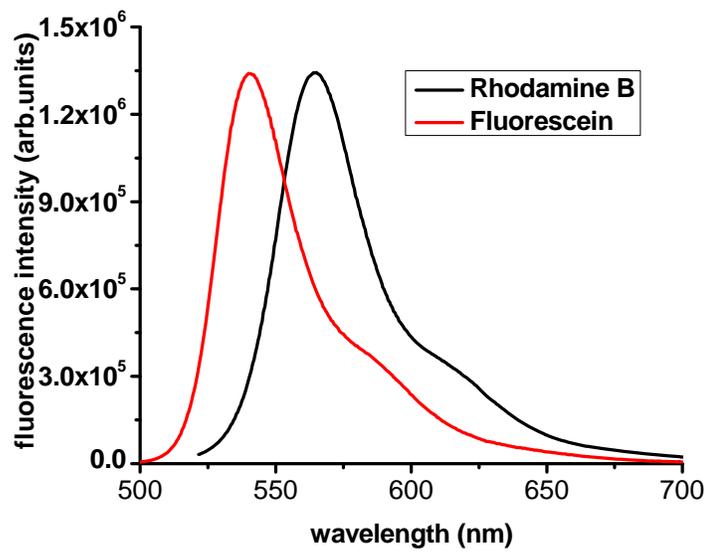

(b)

**Fig. 3**

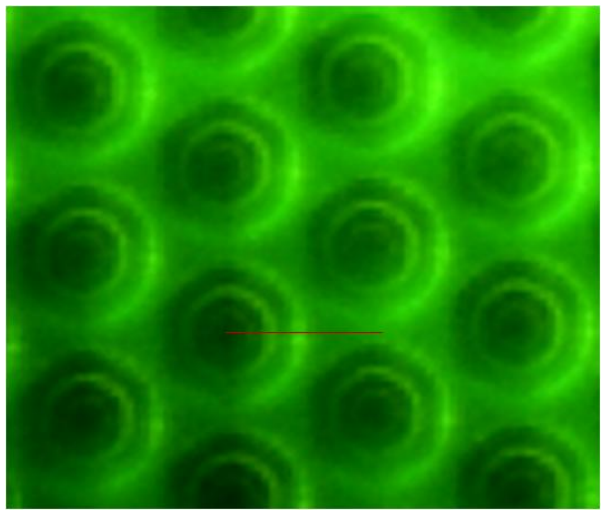
(a)

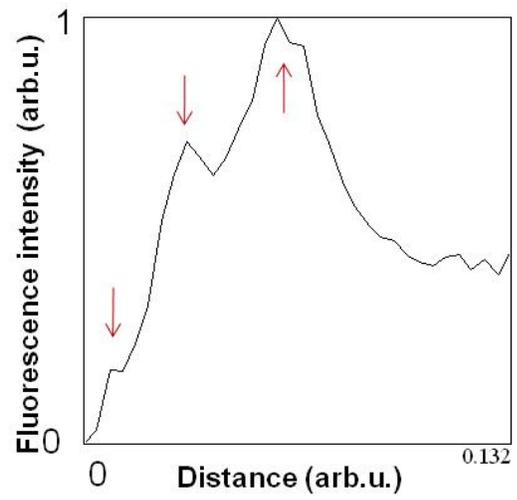
(b)

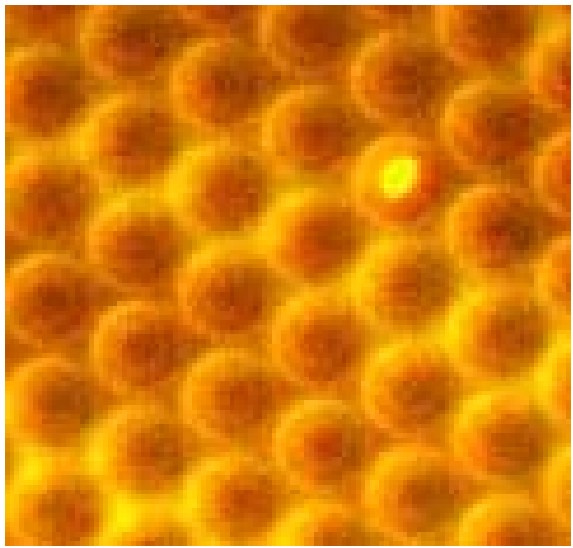
(c)

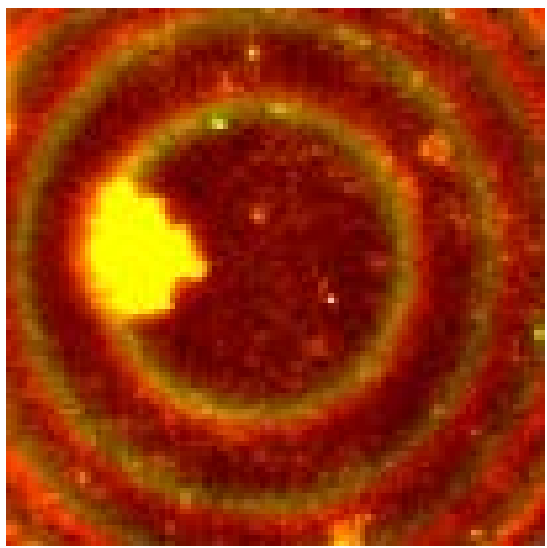
(d)

**Fig. 4**

15